\documentclass[10pt,conference]{IEEEtran}
\IEEEoverridecommandlockouts

\usepackage{cite}
\usepackage{amsmath,amssymb,amsfonts}
\usepackage{graphicx}
\usepackage{textcomp}
\def\BibTeX{{\rm B\kern-.05em{\sc i\kern-.025em b}\kern-.08em
    T\kern-.1667em\lower.7ex\hbox{E}\kern-.125emX}}
\usepackage[utf8]{inputenc}
\usepackage{amsmath}
\usepackage{amssymb}
\usepackage{amsfonts}

\usepackage{amsmath}
\usepackage{color}
\usepackage{amssymb,amsmath,latexsym,amsfonts,amsthm}
\usepackage{mathtools}

\usepackage{algorithm}
\usepackage{algpseudocode}

\usepackage{makecell}

\usepackage{subcaption}
\usepackage{float}
\usepackage{xcolor}
\usepackage{multirow}
\usepackage{booktabs}
\usepackage{mathrsfs}  

\usepackage{float}
\usepackage{upgreek}

\usepackage{multicol}

\usepackage{caption}
\usepackage{subcaption}

\def\BibTeX{{\rm B\kern-.05em{\sc i\kern-.025em b}\kern-.08em
    T\kern-.1667em\lower.7ex\hbox{E}\kern-.125emX}}
\begin{document}



\title{Mitigating Attacks on Artificial Intelligence-based Spectrum Sensing for Cellular Network Signals}


\author{\IEEEauthorblockN{Ferhat Ozgur Catak}
\IEEEauthorblockA{\textit{Dept.
of Electrical Eng. \& Computer Science} \\
\textit{University of Stavanger}\\
Rogaland, Norway \\
f.ozgur.catak@uis.no}
\and
\IEEEauthorblockN{Murat Kuzlu}
\IEEEauthorblockA{\textit{Dept.
of Engineering Technology} \\
\textit{Old Dominion University}\\
Norfolk, VA, USA \\
mkuzlu@odu.edu} 
\and
\IEEEauthorblockN{Salih Sarp}
\IEEEauthorblockA{\textit{Dept. of Electrical and Computer Engineering} \\
\textit{Virginia Commonwealth University}\\
Richmond, VA, USA \\
sarps@vcu.edu}
\and
\IEEEauthorblockN{Evren Catak}
\IEEEauthorblockA{\textit{Independent Researcher} \\
Stavanger, Norway \\
evren.catak@ieee.org}
\and
\IEEEauthorblockN{Umit Cali}
\IEEEauthorblockA{\textit{Dept. of Electric Power Engineering} \\
\textit{Norwegian University of Science and Technology}\\
Trondheim, Norway \\
umit.cali@ntnu.no}
}

\maketitle




\vspace{0.1cm}
\begin{abstract}
Cellular networks (LTE, 5G, and beyond) are dramatically growing with high demand from consumers and more promising than the other wireless networks with advanced telecommunication technologies. The main goal of these networks is to connect  billions of devices, systems, and users with high-speed data transmission, high cell capacity, and low latency, as well as to support a wide range of new applications, such as virtual reality, metaverse, telehealth, online education, autonomous and flying vehicles, advanced manufacturing, and many more.  To achieve these goals, spectrum sensing has been paid more attention, along with new approaches using artificial intelligence (AI) methods for spectrum management in cellular networks. This paper provides a vulnerability analysis of spectrum sensing approaches using AI-based semantic segmentation models for identifying cellular network signals under adversarial attacks with and without defensive distillation methods. 
The results showed that mitigation methods can significantly reduce the vulnerabilities of  AI-based spectrum sensing models against adversarial attacks.
\end{abstract}

\begin{IEEEkeywords}
Adversarial machine learning, artificial intelligence, spectrum sensing, cellular networks
\end{IEEEkeywords}

\section{Introduction}
Cellular networks have experienced substantial attention due to the enormous offerings, such as high-speed data transmission, high cell capacity, and low latency, to support a wide range of new applications. New application areas, i.e.,  virtual reality, metaverse, telehealth services, online education, autonomous and flying vehicles, and many more, require high data rate transmission with low latency. Next-generation cellular networks can easily meet the high demand from the users with advanced communication and computing technologies, i.e., multiple-input multiple-output (MIMO), artificial intelligence (AI), and edge computing \cite{chataut2020massive}. However, the transmission frequency spectrum is a limited resource  and still one of the essential limitations in the advancement of the wireless communication field. Therefore, the radio spectrum is in high demand and has been divided among many services \cite{hu2018full}. That's why spectrum allocation and sharing receive considerable attention for utilizing the limited spectrum fully. Thus, spectrum sensing is one of the research fields to tackle the drawbacks of the limited frequency spectrum \cite{gupta2019progression}.\\
Prior studies propose various techniques for complicated spectrum sensing problems in the literature. Cognitive Radio (CR) techniques are adopted for spectrum sensing to ease the scarcity of spectrum resources \cite{nasser2021spectrum}. Conventional methods often need to full or limited prior information about primary users. In addition, CR sensing methods are susceptible to noise uncertainty, leading to lower sensing precision \cite{obite2021overview}. Besides traditional CR methods, new ways to monitor wireless spectrum are developed with the introduction of AI into the 5G and beyond wireless communication. AI models could untangle complicated feature extraction and learning tasks. The study in \cite{du2020spectrum} proposes a spectrum sensing model consisting of an information geometry and a deep learning (DL) classifier. Their model outperforms traditional models by reaching better sensing precision. The authors in \cite{gao2019deep} present a DL-based spectrum sensing model that utilizes structural information of the modulated signal without any prior knowledge of channel state. The proposed model also yields better performance in comparison with traditional sensing techniques. Although the adoption of AI provides simpler but powerful solutions to complex problems, the security aspect of these implementations has not been explored deeply \cite{sarp2021use}. AI-powered models are prone to attacks that can cause security \cite{CATAK2022101626,9527756}
and privacy violations \cite{wibawa2022homomorphic}. The adversaries could also poison pre-trained AI models to take the place of legitimate models \cite{kuzlu2021role}. To overcome these security threads, adversarial attacks and training methods should be developed and adopted. 

This study presents vulnerability analysis of AI-based spectrum sensing, i.e., CNN-based semantic segmentation model, to identify cellular network signals for spectrum monitoring. This analysis also includes the comparison of adversarial attacks along with a mitigation method to evaluate the models' robustness.


\section{Preliminaries}
\subsection{Spectrum Sensing}
In wireless communication, the radio spectrum is one of the key resources. However, it is limited, and not fully used due to several reasons, such as region-based regulations or technical difficulties. Fortunately, the existing radio spectrum has more been allocated to consumers along with the high demand in the last decades. A large part of the existing radio spectrum is also licensed and allocated for service providers, such as cellular communication, television
broadcasting, radio, military applications, etc. According to the report released by the Federal Communications Commission (FCC),  some parts of the spectrum are barely utilized, e.g., the spectrum utilization in the 0–6 GHz band is between 15\%
and 85\% \cite{ma2009signal}. Therefore, FCC recommends that free bands can be used by the secondary user(s) until they do not cause any issues with primary users’ communication. The spectrum exploring for free bands is called ``Spectrum
Sensing". It is the process of periodically checking a specific frequency band to identify the occupied frequency bands for users or services. Spectrum sensing is also a fundamental problem for cognitive radio (CR), which has become a novel form of wireless communications. CR is an intelligent  wireless communication method using three main steps, (1) sensing
the outside electromagnetic environment, (2) learning from the surroundings, and (3) adapting the internal states and operating parameters, such as transmit power, carrier frequency,  modulation strategy, etc. 

In the literature, there are satisfactory studies in spectrum sensing and related topics. For example, the study \cite{zeng2010review} provides a comprehensive analysis of spectrum sensing techniques for CR, namely the optimal likelihood ratio test, energy detection, matched filtering detection, cyclostationary detection, eigenvalue-based sensing, joint space-time sensing, and robust sensing methods. It also indicated that the source signal and the propagation channel are two important factors for the spectrum sensing along with the selected methods to provide better performance.
Although there have been many methods proposed
for spectrum sensing, they are still suffering due to uncertainty of the nature of the communication channel and some issues, including narrow-band noise, spurious signal and interference, fixed point realization, wide-band sensing, and complexity. Fortunately, AI methods have been started to use in communication systems, especially wireless ones, for spectrum sensing to  overcome the challenges existing spectrum sensing techniques suffer.\\

\subsection{Adversarial Machine Learning Attacks}
Most machine learning models are highly vulnerable to Adversarial Machine Learning (AML) attacks. The adversarial attack is one type of widely used cyberattacks to poison a model during the training phase. It provides misleading data, i.e., manipulated input with a slight difference or adversarial examples (AEs), to reduce the model performance in terms of accuracy. 
An AE cannot usually be noticeable by a human; however, it can cause the misclassification and misdirection of the model. 
This study focuses on the most popular three adversarial attacks, (1) Fast Gradient Sign Method (FGSM), (2) Basic Iterative Method (BIM), and (3) Projected Gradient Descent (PGD).

\begin{itemize}
  \item FGSM is the most popular and one of the simplest AML attacks to generate adversarial inputs, which was first introduced by Goodfellow \textit{et al}. in \cite{goodfellow2014explaining}. This attack uses gradient information, i.e., the partial derivative of the model output for the input data, to determine the direction of the perturbations.
  FGSM computes the gradients of a loss function (e.g., mean-squared error or categorical cross-entropy) and generates adversarial examples by adding the gradient sign to the input data. The general formula of the FGSM is defined as follow:

\begin{equation}\label{FGSM}
x_{adv} = x_{0} + \epsilon \cdot \text{sign}(\nabla_{x} J(\theta, x_{0}, y_{0}))
\end{equation}
where $x_{adv}$: the adversarial example, $x_{0}$: the legitimate input data, $y_{0}$: the true label of $x_{0}$, $J(\theta, x_{0}, y_{0})$: the loss function, $\nabla_{x} J(\theta, x_{0}, y_{0})$: the gradient of the loss function. 

\item BIM is an improved or an iterative version of the FGSM, which was introduced by Kurakin \textit{et al}. in \cite{kurakin2016adversarial}. Instead of taking one large step like FGSM, a BIM attack generates adversarial examples using an iterative approach by applying FGSM iteratively many times with small steps, i.e., $\alpha$. This process continues by misleading the model or reaching the allowed maximum perturbation. 
A BIM attack can be defined as follow:
\begin{equation}
x_{t+1} = \textit{Clip}_{x, \epsilon} \{x_{t} + \alpha \cdot \text{sign}(\nabla_{x} J(\theta, x_{t}, y_{0}))\}
\end{equation}
where: \textit{Clip} is the function to limit the maximum perturbance, $t$ is the iteration index, $\alpha$ is the step size which is set to be 1 to minimize the number of iterations.



\item PGD is similar to the BIM attack. However, its capability is more than both FGSM and BIM, and stronger in terms of first-order attack, and a bit slower as expected. Instead of initializing to the original point like BIM, it initializes the search for an adversarial example at a random point within the allowed norm ball to find adversarial examples \cite{kannan2018adversarial}. The formula of the PGD attack is defined as follows:

\begin{equation}
x_{t+1} = \text{Clip}_{x, \epsilon} \{x_{t} + \alpha \cdot \text{sign}(\nabla_{x} J(\theta, \text{Clip}_{x, \epsilon} \{x_{t}\}, y_{0}))\}
\end{equation}
\end{itemize}

\subsection{Convolutional Neural Networks}
\label{sec:cnn}

The CNN is one of the key computer vision technology that facilitate feature extraction with its ability of capturing spatial and temporal dependencies by utilizing various filters \cite{5537907}. Images consist of pixels that can be treated as two-dimension matrices, $\mathbf{x}$. A convolution operation between the image and a filter, $\mathbf{W}$, can be defined as :  

\begin{equation}
\label{eq:conv}
\mathbf{y} = \mathbf{W} \ast \mathbf{x} = \sum_{i=1}^{W} \sum_{j=1}^{H} \mathbf{W}_{i,j} \mathbf{x}_{i-s,j-s},
\end{equation}
where the width and height of the image $\mathbf{x}$ are defined as $W$ and $H$ with the number of strides, $s$.


Several layers are used in a CNN, such as convolution, pooling, and fully connected layers. The combination of these layers is used to create state-of-the-art computer vision models with varying parameters. The convolutional layer benefits from different filters to extract divergent features. The pooling layer generalizes the extracted features and lowers the size of the output by either pooling the max or mean values of a pixel group. In the last part of CNN, fully connected layers are used to combine all the features, and a special softmax layer is deployed to convert numbers to probabilities which enhances the classification tasks.

\subsection{Knowledge Distillation}
Knowledge distillation (KD) \cite{hinton2015distilling} is utilized to transfer knowledge from a teacher to student model by lowering the output entropy. Hard labels will guide in making of teacher model. However, soft labels (probabilistic labels) are employed to train student network which increase the capacity and efficiency. The general KD technique is defined as:

 
 \begin{equation}
 \mathcal{L}_{KD} = \mathcal{L}_{CE}(f_{s}(x), y) + \lambda \mathcal{L}_{KL}(f_{s}(x), f_{t}(x))
 \end{equation}
  where the cross entropy loss and Kullback–Leibler divergence are written as $\mathcal{L}_{CE}$, and $\mathcal{L}_{KL}$, respectively. The softmax outputs of the teacher and student models are $f_{t}(\cdot)$ and $f_{s}(\cdot)$, respectively with the $\lambda$ as the weighting parameter. The Kullback–Leibler divergence is given as:

 
 \begin{equation}
 \mathcal{L}_{CE}(f_{s}(x), y) = - \sum_{i=1}^{c} y_{i} \log(f_{s}(x)_{i}).
 \end{equation}
 
 The $c$ and $y_{i}$ are the number of classes for the $i$th element of the label vector. The $\mathcal{L}_{KL}$ is specified as:
 
 \begin{equation}
 \mathcal{L}_{KL}(f_{s}(x), f_{t}(x)) = - \sum_{i=1}^{c} f_{s}(x)_{i} \log(\frac{f_{t}(x)_{i}}{f_{s}(x)_{i}}).
 \end{equation}


Defensive distillation (DD) is a way to defend the CNN model against adversaries by utilizing the KD idea \cite{papernot2016distillation}. This method prevent finding AEs for the distilled network, even if the teacher network is compromised. The formula for the DD is given as:
 \begin{equation}
 \mathcal{L}_{DD} = \mathcal{L}_{CE}(f_{s}(x), y) + \lambda \mathcal{L}_{KL}(f_{s}(x), f_{t}(x)).
 \end{equation}

\begin{figure*}[htbp!]
    \centering
    \includegraphics[width=0.8\linewidth]{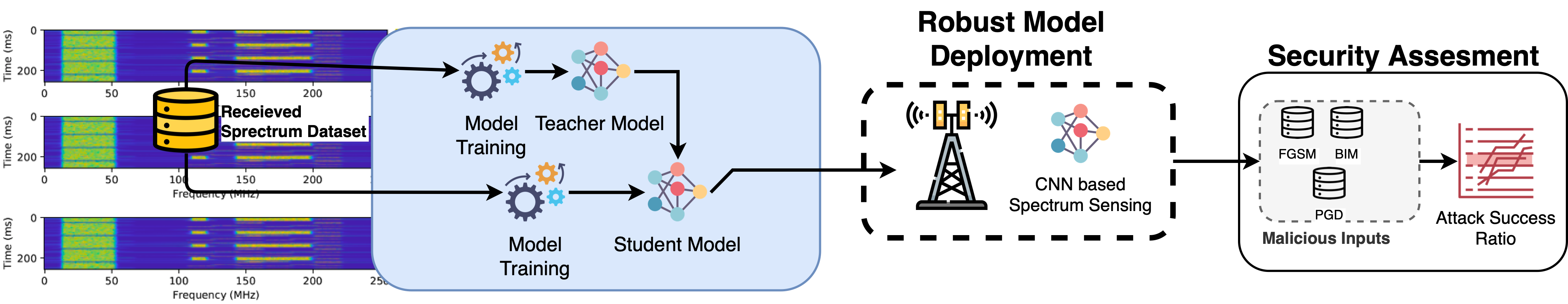}
    \caption{Overview of knowledge distillation.}
    \label{fig:knowledge_distillation_figure}
\end{figure*}

Figure \ref{fig:knowledge_distillation_figure} shows the overall steps for this technique. The knowledge distillation process consists of two steps: (1) training the teacher model and (2) distilling the knowledge from the teacher to the student. According to the figure, the distillation can be performed using the teacher model's output probabilities, the teacher model's activations, or the intermediate representations of the teacher model. Finally, the robust student model would be deployed to the base stations for spectrum sensing.

\section{Dataset Description, Spectrum Sensing Scenario, and Performance Metrics}\label{sec:dataset}
\subsection{Dataset Description}


In this study, the synthesized dataset is utilized and generated by using two MATLAB toolboxes, i.e., 5G signals in 5G Toolbox and LTE signals in LTE Toolbox \cite{matlab_example_ss}. Then, the generated dataset, including LTE and 5G  signals, is divided into the training and testing datasets with a ratio of 80/20\%, respectively. Each frame of 40 ms is randomly shifted in the frequency domain. It is assumed that LTE and 5G signals are in the selected band range, and the network performance is evaluated based on distinct random bands. In this example, the sampling rate is 61.44 MHz, which is adequate to process LTE and 5G signals. Respective 256 by 256 RGB spectrogram images are generated from complex baseband signals using an FFT (Fast Fourier Transform) length of 4096. The class imbalance occurs as there is a vast noise filling the background and larger bandwidth of 5G than LTE. It is rectified using class weighting to mitigate training bias towards dominant classes.



\subsection{Spectrum Sensing Scenario }
Spectrum sensing implies the detection of white spaces and the characterization of the frequency spectrum. The CNN-based model identifies the cellular signal(s). Parameters of both LTE and  5G signal generation are indicated in Table \ref{tab:datasetParameter}. In the table, SCS presents the sub-carrier spacing, SSB presents the single sideband,  while BW is bandwidth. CNN-based spectrum sensing model identifies the type of the cellular network signal, i.e., 4G, 5G, or noise, based on the created spectrogram images from complex baseband signals. Each pixel of the image is labeled as one of the type signals, i.e., LTE,  5G or noise, for the class name mapping. The optimized neural network model is trained once and does not need to be further trained and optimized for weights. 

\begin{table}[!htbp]
\centering
\caption{The parameters of LTE and 5G signals}
\label{tab:datasetParameter}
\begin{tabular}{|l|c|c|}
\hline
\textbf{Channel Parameter}& {\textbf{Value }} & Unit \\
 \hline 
5G BW & [10 15 20 25 30 40 50 ] & MHz\\
 \hline 
 5G SCS & [15 30 ] & kHz\\
 \hline 
 5G SSB Block Pattern & ["Case A" "Case B'] & - \\
 \hline 
  5G Period & [20] & ms\\
 \hline 
  LTE Reference Channel & ["R.2", "R.6", "R.8", "R.9"] & - \\
 \hline 
  LTE BW & [10 5 15 20 ] & MHz\\
 \hline 
  LTE Dublex Mode & FDD & -\\
 \hline 
  Channel SNR & [40 50 100] & dB\\
  \hline 
  Channel Doppler & [0 10 500] & Hz\\
  
 \hline 
\end{tabular}
\label{tab:dataset}
\end{table}


\subsection{Performance Metrics}
The performance of the models is evaluated and compared with  the following metrics.

\emph{\textit{Accuracy}}: It is the correctly classified pixel percentage. 

\emph{\textit{Recall}}:  Recall or sensitivity metric gives the completeness of positively identified pixels compared to the number of actual positive pixels. 

\emph{\textit{False Positive Rate (FPR)}}: This rate is the ratio between wrongly identified negative pixels over the total number of actual negative pixels.   

\emph{\textit{Precision}}:  It measures the purity of the positively identified pixels in proportion to the number of actual positive pixels.

\emph{\textit{Specificity}}:  It is also called true negativity rate, and measures the correctly identified negative pixel ratio over the total number of actual negative pixels.

\emph{\textit{F-Score}}:  It is the harmonic mean of the precision for a better evaluation.

\emph{\textit{Intersection of Union (IoU) }}: It is the percentage of overlap between the target mask and predicted output.



\section{Experiments}\label{sec:experiments}
\begin{figure}[htbp!]
    \centering
     \includegraphics[width=0.9\linewidth]{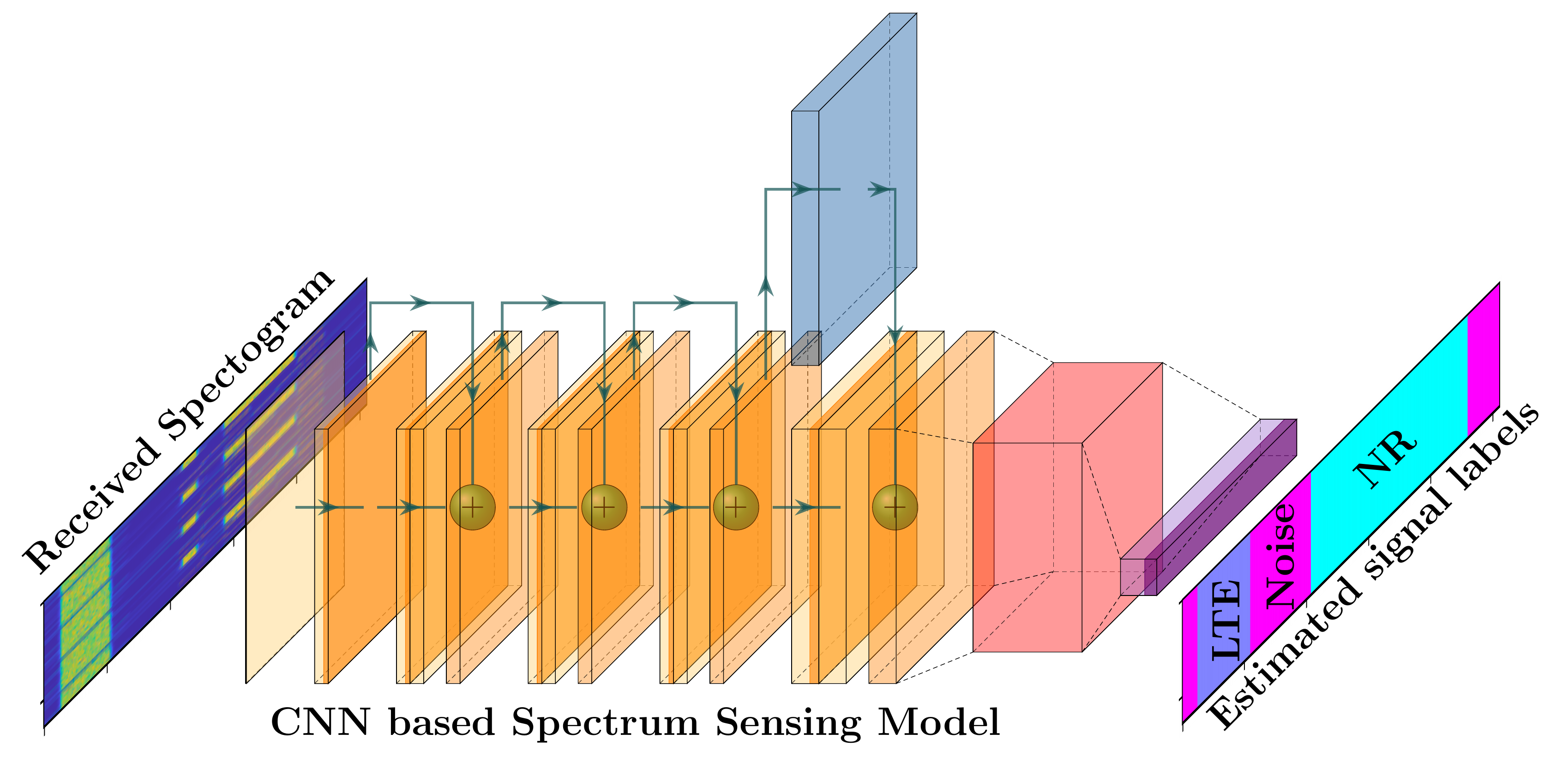}
    \caption{CNN model architecture}
    \label{fig:model}
\end{figure}

\begin{figure}[!htbp]
     \centering
     \begin{subfigure}[b]{0.9\linewidth}
         \centering
         \includegraphics[width=\linewidth]{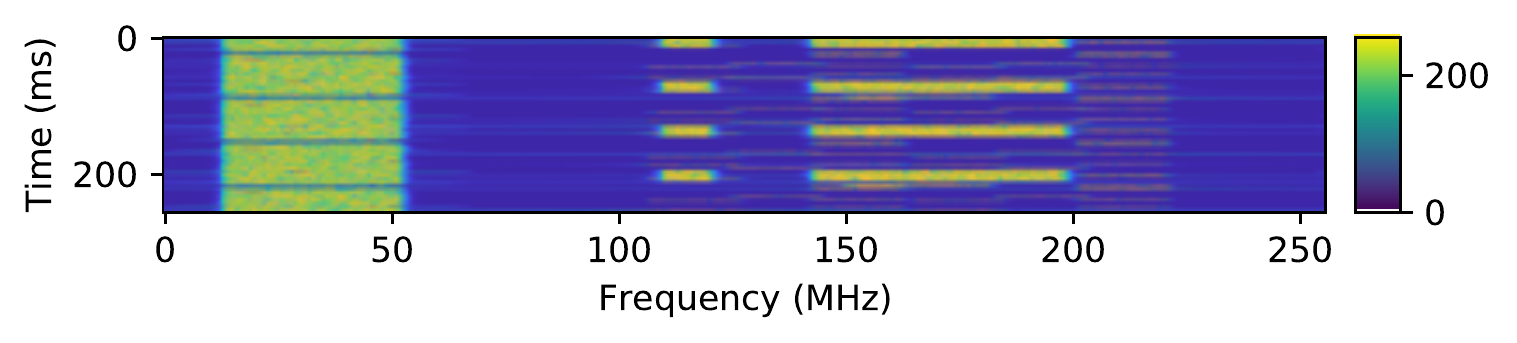}
         \caption{Received spectogram}
         \label{fig:recv}
     \end{subfigure}
     \hfill
     \begin{subfigure}[b]{0.9\linewidth}
         \centering
         \includegraphics[width=\linewidth]{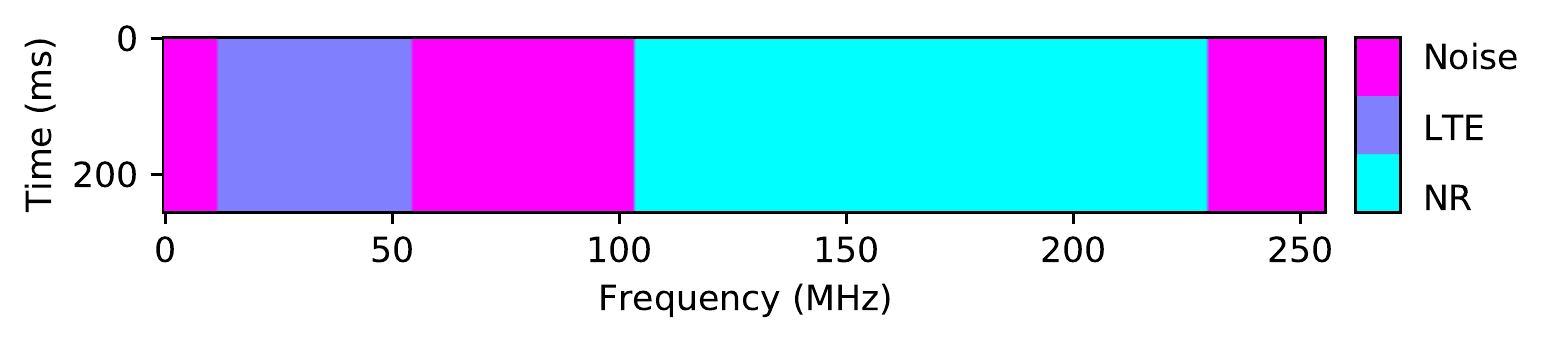}
         \caption{True signal labels}
         \label{fig:true_signal}
     \end{subfigure}
     \hfill
     \begin{subfigure}[b]{0.9\linewidth}
         \centering
         \includegraphics[width=\linewidth]{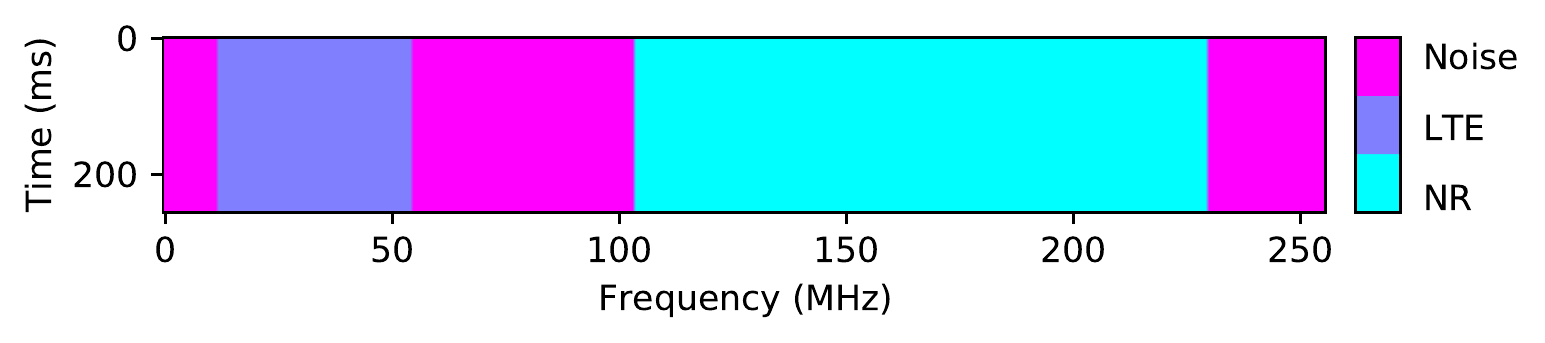}
         \caption{Estimated signal labels}
         \label{fig:estimated_signal}
     \end{subfigure}
     \hfill
     \begin{subfigure}[b]{0.8\linewidth}
         \centering
         \includegraphics[width=\linewidth]{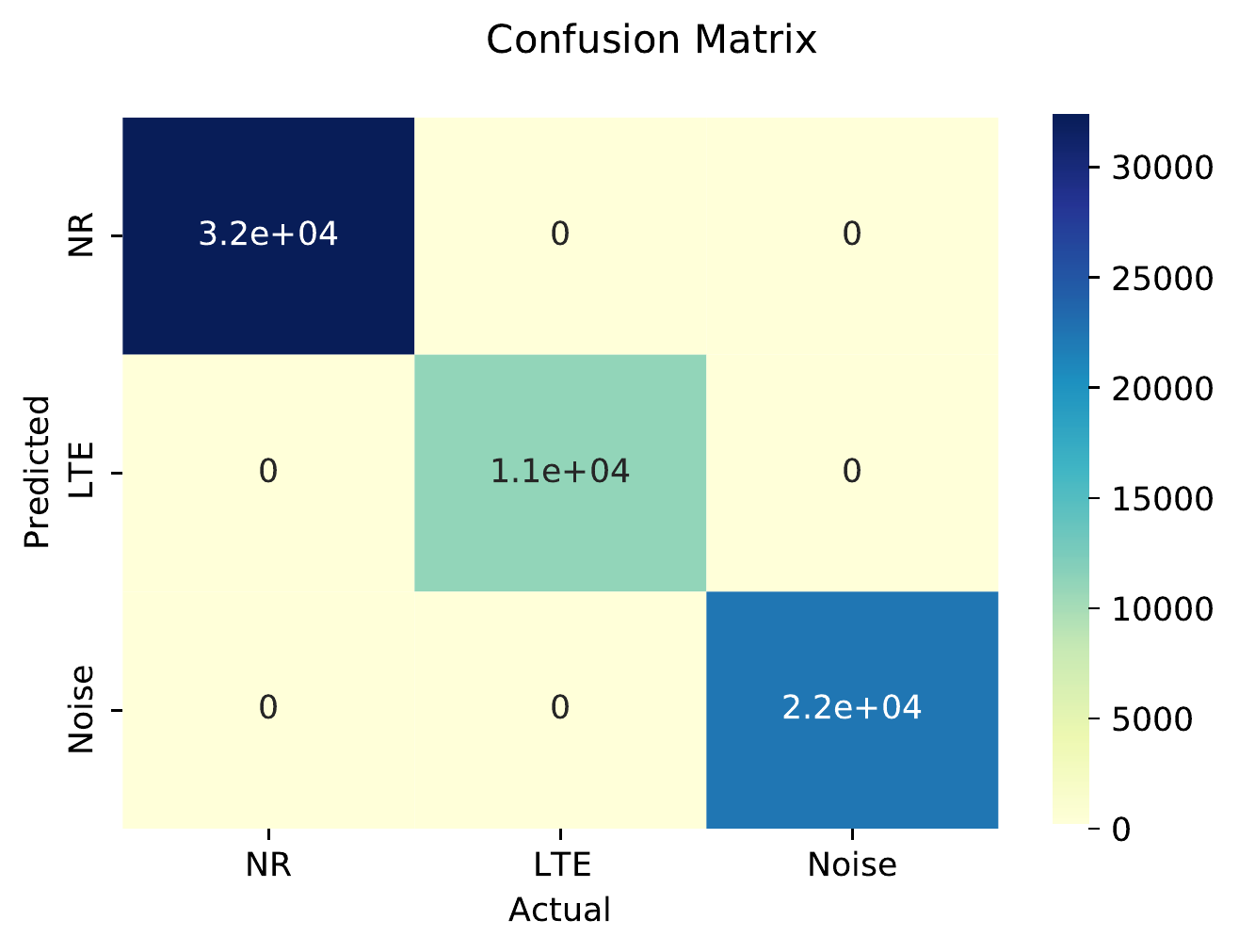}
         \caption{Confusion matrix of the true and estimated signal}
         \label{fig:cm}
     \end{subfigure}
        \caption{Example sample to identify LTE and 5G signals in the spectrogram}
        \label{fig:example_dataset_instance}
\end{figure}

The proposed approach is implemented in the Python programming language. The CNN-based segmentation method, i.e., semantic segmentation, is based on RESNET50 and  re-trained using the dataset generated by using the LTE and 5G toolboxes of the MATLAB R2022a. The \textit{Rmsprop} optimizer is used for training the model. The cross-entropy is used as the loss function. The training and validation data split ratio is 80/20\%, respectively. The base model is trained in two phases using the proposed approach (1) designed to train the base model and (2) designed to defend the base model against the AML attack. The base model is trained with the cross-entropy loss function. The proposed approach is designed to train the distilled network with the loss of a combination of cross-entropy and KL divergence, which is called defensive distillation. The model architecture of the spectrum sensing model, as shown in Figure \ref{fig:model}. In the Figure \ref{fig:model} and \ref{fig:example_dataset_instance}, NR (New Radio) represents 5G or 5G NR. 

Figure \ref{fig:example_dataset_instance} shows one of the example instances from the dataset. Figure \ref{fig:recv} is the received spectrogram (i.e., the input for the CNN model), Figure \ref{fig:true_signal} is the true signal labels (i.e., the real output of the input), Figure \ref{fig:estimated_signal} is the prediction of the CNN-based spectrum sensing model with the received spectrogram, and Figure \ref{fig:cm} is the confusion matrix of the prediction.

Table \ref{tab:init_res} shows the initial prediction performance results of the CNN-based spectrum sensing model without attacks with different metrics for LTE, 5G, and noise.

\begin{table}[!htbp]
    \centering
    \caption{Initial prediction performance results}
    \label{tab:init_res}
\begin{tabular}{lrrrr}
\hline
{} &   \textbf{Average} &            \textbf{5G} &           \textbf{LTE} &         \textbf{Noise} \\
\hline
\textbf{Accuracy}           &  0.967200 &      0.976256 &      0.978675 &      0.979470 \\
\textbf{Recall}             &  0.967473 &      0.944062 &      0.990233 &      0.968123 \\
\textbf{Precision}          &  0.954690 &      0.971999 &      0.945672 &      0.946400 \\
\textbf{Specificity}        &  0.984998 &      0.998104 &      0.976199 &      0.980692 \\
\textbf{F-Score}            &  0.953455 &      0.951235 &      0.959088 &      0.950043 \\
\textbf{FPR}                &  0.015002 &      0.001896 &      0.023801 &      0.019308 \\
\textbf{IoU}                &  0.934285 &      0.941315 &      0.937662 &      0.923879 \\
\hline
\end{tabular}
\end{table}

For experiments, the second step is to attack the CNN model with  FGSM, BIM, and PGD attacks. The $\epsilon$ values of the FGSM, BIM, and PGD attacks are selected from 13, 26, 38, 51, 64, 76, 89, 102, 115, to 128. The number of iterations involved in the BIM and PGD attacks is 2000. 

Table \ref{tab:undef} - \ref{tab:def} together show the prediction performance results  of the defended and undefended CNN-based spectrum sensing models against the attacked for three $\epsilon$ values, i.e., 13, 64, 128, along with different metrics for LTE, 5G, and noise. According to the tables, the defended CNN model has a higher prediction performance when compared to the undefended model. 

\begin{table}[!htbp]
    \centering
    \caption{Undefended}
    \label{tab:undef}
    \begin{tabular}{|c|l|c|c|c|}
    \hline
        $\epsilon$ & \textbf{Metrics} & \textbf{FGSM} & \textbf{BIM} & \textbf{PGD}  \\ \hline \hline
        \multirow{7}{*}{\textbf{13}} &   \textbf{Accuracy}   &0.999485 &0.999107 &0.999485  \\
                            &    \textbf{Recall}     &0.999418 &0.998906 &0.999418  \\
                            &    \textbf{Precision}  &0.999316 &0.998817 &0.999316  \\
                            &    \textbf{Specificity}&0.999771 &0.999604 &0.999771  \\
                            &    \textbf{F-Score}    &0.999365 &0.998857 &0.999365  \\
                            &    \textbf{FPR}        &0.000229 &0.000396 &0.000229  \\
                            &    \textbf{IoU}        &0.998733 &0.997723 &0.998733  \\ \hline \hline
        \multirow{7}{*}{\textbf{64}} &   \textbf{Accuracy}   &0.999367 &0.859735 &0.920404  \\
                            &    \textbf{Recall}     &0.999514 &0.862166 &0.908336  \\
                            &    \textbf{Precision}  &0.998738 &0.837363 &0.905915  \\
                            &    \textbf{Specificity}&0.999733 &0.940058 &0.965371  \\
                            &    \textbf{F-Score}    &0.999111 &0.813058 &0.883668  \\
                            &    \textbf{FPR}        &0.000267 &0.059942 &0.034629  \\
                            &    \textbf{IoU}        &0.998253 &0.722727 &0.817422  \\ \hline \hline
        \multirow{7}{*}{\textbf{128}}&   \textbf{Accuracy}    &0.999678 &0.817982 &0.845656 \\
                            &    \textbf{Recall}     &0.999688 &0.825223 &0.863721  \\
                            &    \textbf{Precision}  &0.999568 &0.789182 &0.814875  \\
                            &    \textbf{Specificity}&0.999855 &0.914792 &0.933859  \\
                            &    \textbf{F-Score}    &0.999626 &0.761299 &0.803141  \\
                            &    \textbf{FPR}        &0.000145 &0.085208 &0.066141  \\
                            &    \textbf{IoU}        &0.999256 &0.668107 &0.715293  \\ \hline
    \end{tabular}
\end{table}

\begin{table}[!htbp]
    \centering
    \caption{Defended}
    \label{tab:def}
    \begin{tabular}{|c|l|c|c|c|}
    \hline
        $\epsilon$ & \textbf{Metrics} & \textbf{FGSM} & \textbf{BIM} & \textbf{PGD}  \\ \hline \hline
        \multirow{7}{*}{\textbf{13}} &   \textbf{Accuracy}   &0.999485 & 0.999107 & 0.999485   \\
                            &    \textbf{Recall}     &0.999418 & 0.998906 & 0.999418  \\
                            &    \textbf{Precision}  &0.999316 & 0.998817 & 0.999316  \\
                            &    \textbf{Specificity}&0.999771 & 0.999604 & 0.999771  \\
                            &    \textbf{F-Score}    &0.999365 & 0.998857 & 0.999365  \\
                            &    \textbf{FPR}        &0.000229 & 0.000396 & 0.000229  \\
                            &    \textbf{IoU}        &0.998733 & 0.997723 & 0.998733  \\ \hline \hline
        \multirow{7}{*}{\textbf{64}} &   \textbf{Accuracy}   &0.999559 & 0.938550 & 0.944829  \\
                            &    \textbf{Recall}     &0.999418 & 0.999170 & 0.938213  \\
                            &    \textbf{Precision}  &0.999064 & 0.930787 & 0.941644  \\
                            &    \textbf{Specificity}&0.999811 & 0.972697 & 0.974934  \\
                            &    \textbf{F-Score}    &0.999102 & 0.919045 & 0.930235  \\
                            &    \textbf{FPR}        &0.000189 & 0.027303 & 0.025066  \\
                            &    \textbf{IoU}        &0.998237 & 0.881291 & 0.896488  \\ \hline \hline
        \multirow{7}{*}{\textbf{128}}&   \textbf{Accuracy}    &0.998690 & 0.882667 & 0.904468 \\
                            &    \textbf{Recall}     &0.998900 & 0.865154 & 0.9067030  \\
                            &    \textbf{Precision}  &0.997399 & 0.880294 & 0.910402  \\
                            &    \textbf{Specificity}&0.999467 & 0.934967 & 0.956826  \\
                            &    \textbf{F-Score}    &0.998007 & 0.840553 & 0.884733  \\
                            &    \textbf{FPR}        &0.000533 & 0.065033 & 0.043174  \\
                            &    \textbf{IoU}        &0.996308 & 0.802265 & 0.844832  \\ \hline
    \end{tabular}
\end{table}

\begin{figure}[!tbp]
     \centering
     \begin{subfigure}[b]{0.49\linewidth}
         \centering
         \includegraphics[width=0.9\linewidth]{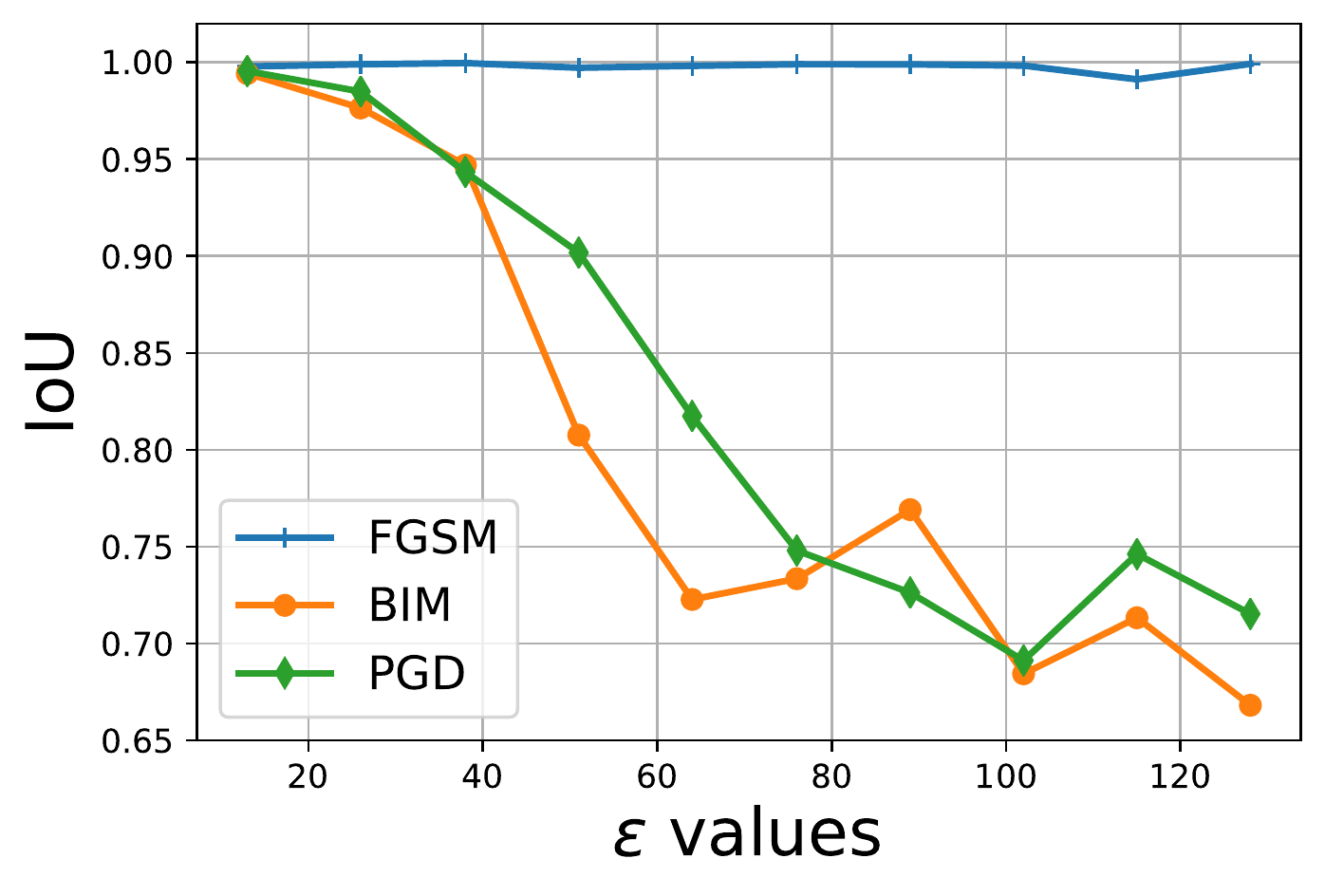}
         \caption{Undefended}
         \label{fig:undefended}
     \end{subfigure}
     \hfill
     \begin{subfigure}[b]{0.49\linewidth}
         \centering
         \includegraphics[width=0.9\linewidth]{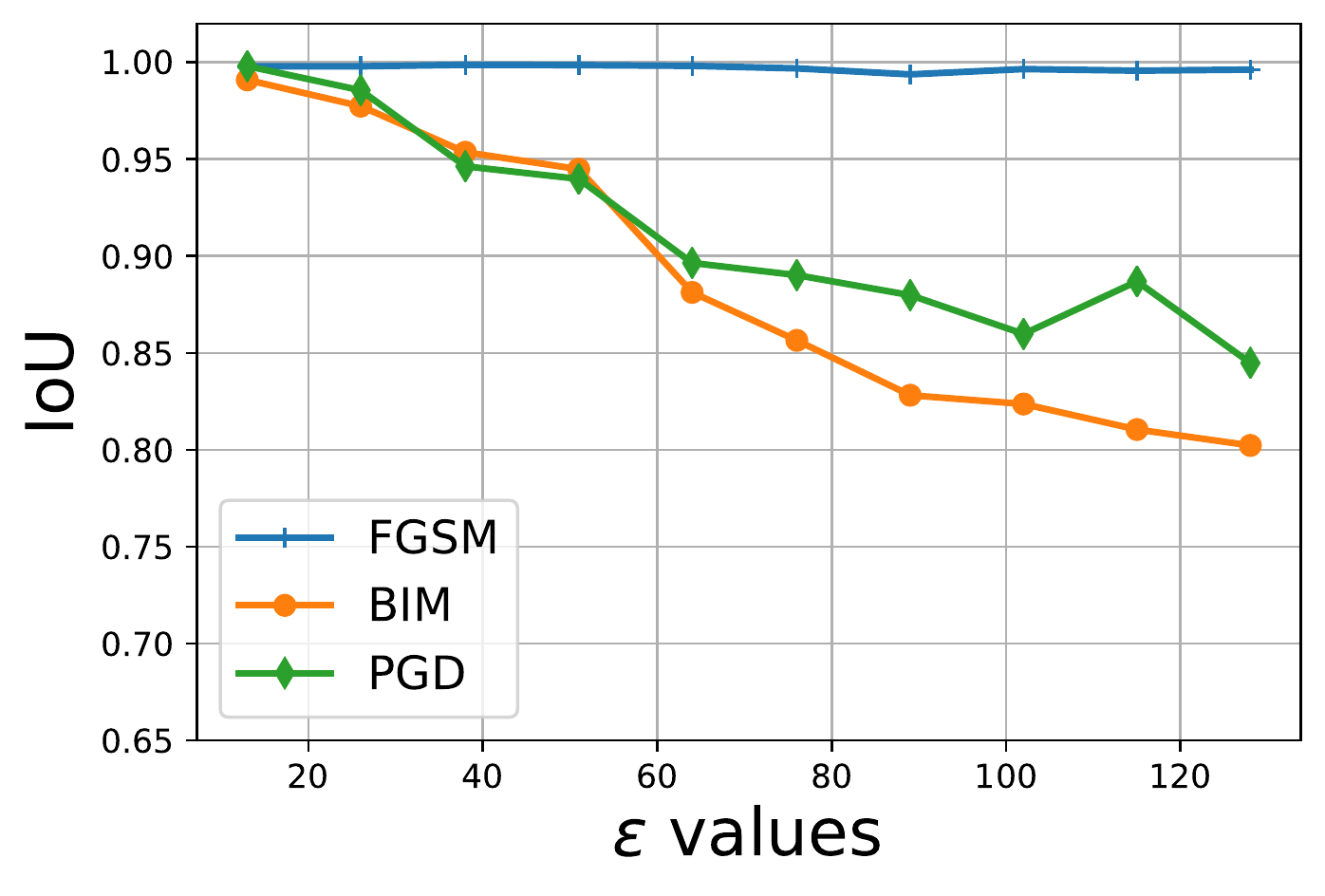}
         \caption{Defensive distillation}
         \label{fig:defended}
     \end{subfigure}
        \caption{IoU metric values of the adversarial attacks 
        }
        \label{fig:iou_res}
\end{figure}

IoU metric values of the defended, i.e., robust student model, and the undefended model, i.e., without mitigation method, are shown in Figure \ref{fig:undefended}  and Figure \ref{fig:defended}, respectively. The X-axis indicates the $\epsilon$ values of the FGSM, BIM, and PGD attacks, while Y-axis indicates the IoU values of the defended and undefended models. As shown in the figures, the defended model has higher IoU metric values when compared to the undefended model.

\begin{figure}[!tbp]
     \centering
     \begin{subfigure}[b]{0.49\linewidth}
         \centering
         \includegraphics[width=0.9\linewidth]{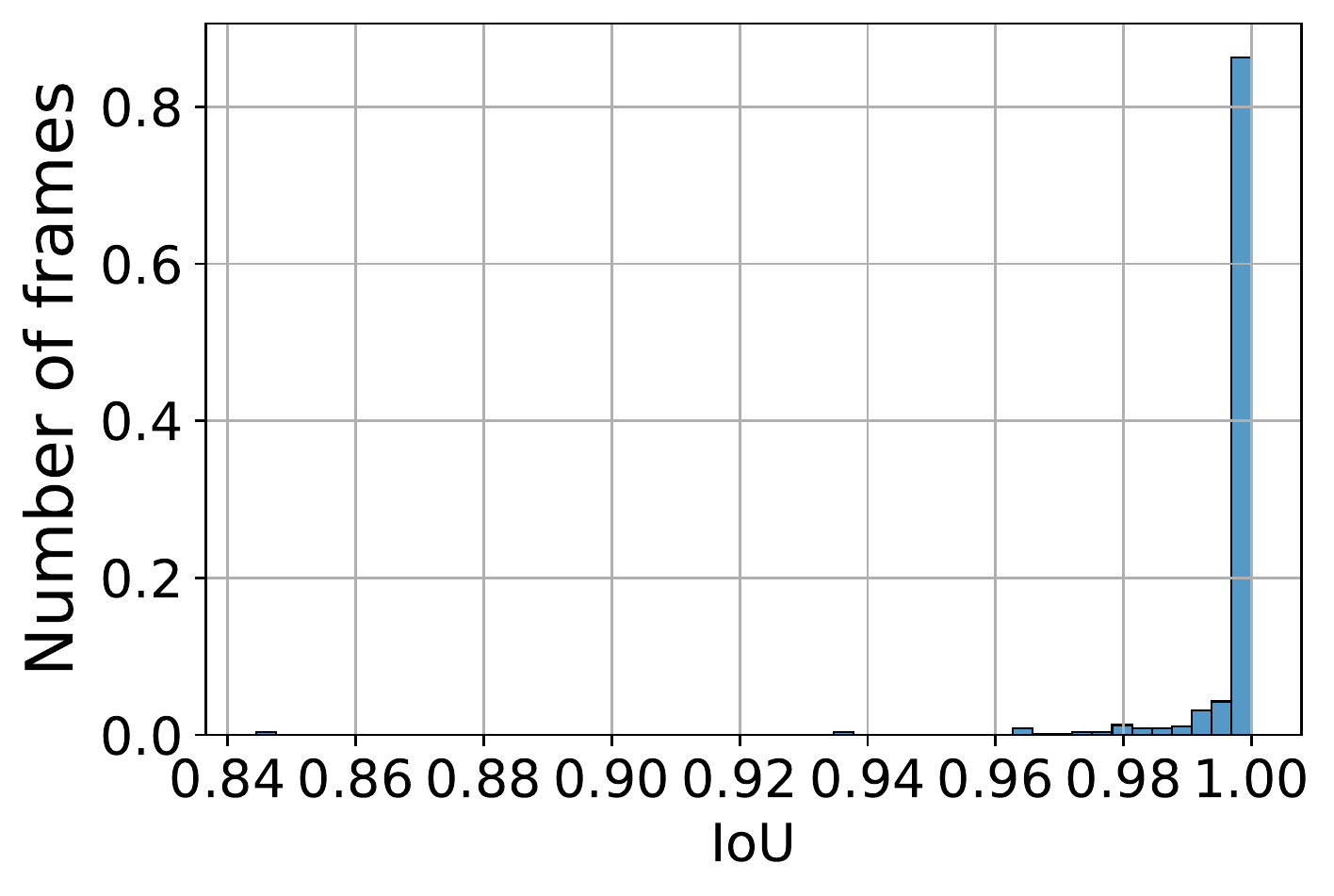}
         \caption{Undefended}
         \label{fig:undefended_iou_fgsm}
     \end{subfigure}
     \hfill
     \begin{subfigure}[b]{0.49\linewidth}
         \centering
         \includegraphics[width=0.9\linewidth]{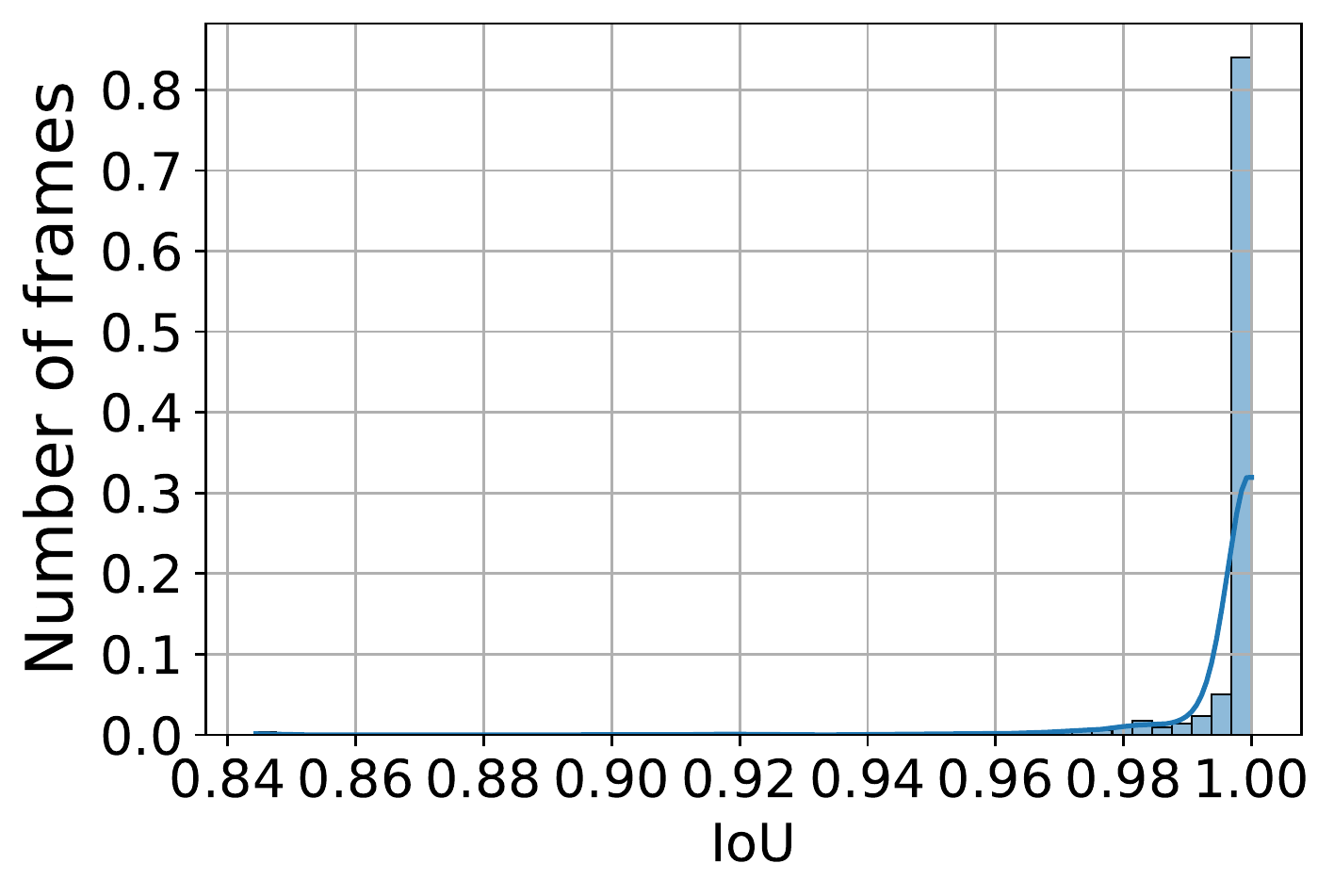}
         \caption{Defensive distillation}
         \label{fig:defended_iou_fgsm}
     \end{subfigure}
    \caption{FGSM}
        \label{fig:fgsm_iou_bar}
\end{figure}

\begin{figure}[!tbp]
     \centering
     \begin{subfigure}[b]{0.49\linewidth}
         \centering
         \includegraphics[width=0.9\linewidth]{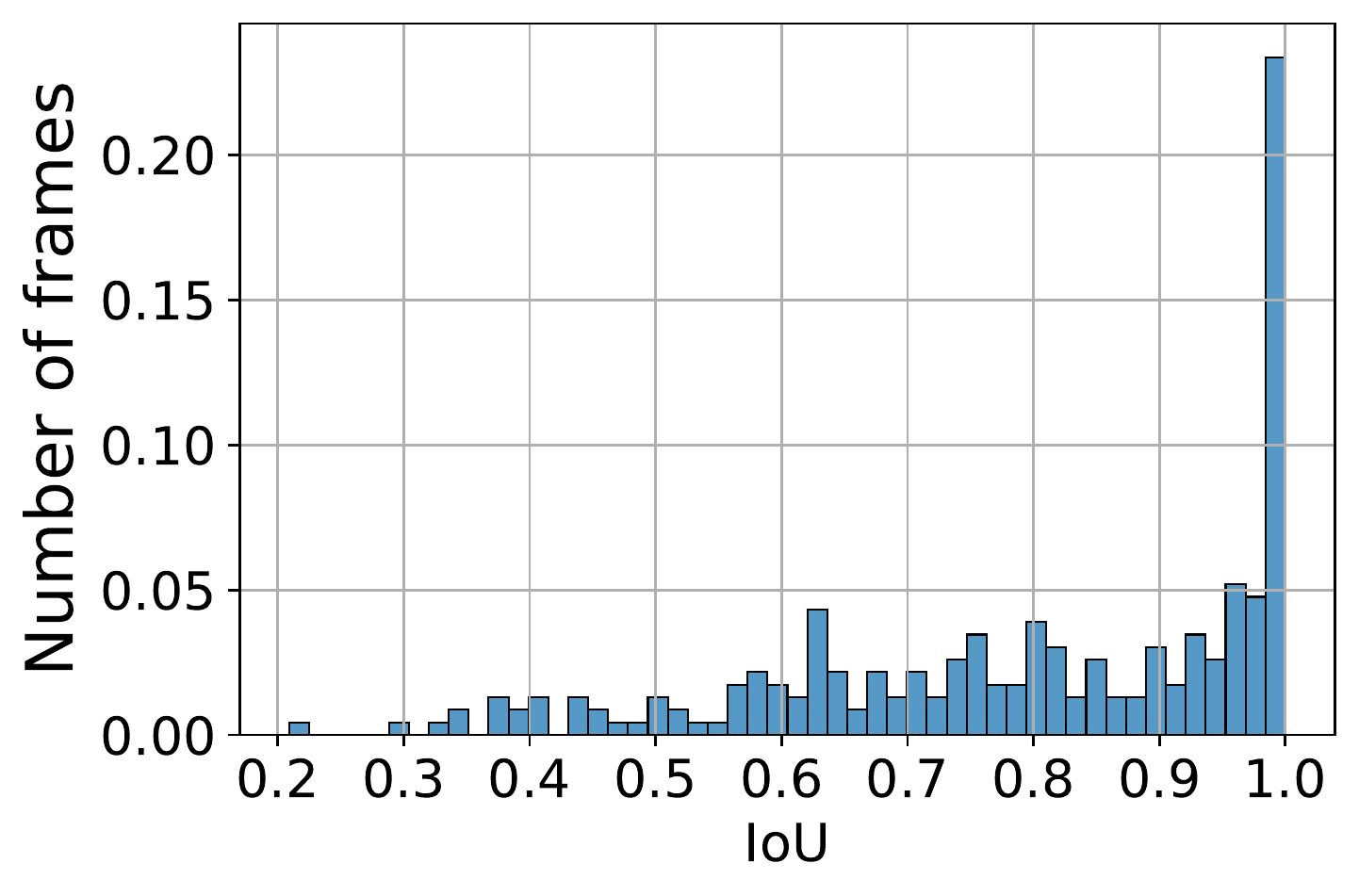}
         \caption{Undefended}
         \label{fig:undefended_iou_bim}
     \end{subfigure}
     \hfill
     \begin{subfigure}[b]{0.49\linewidth}
         \centering
         \includegraphics[width=0.9\linewidth]{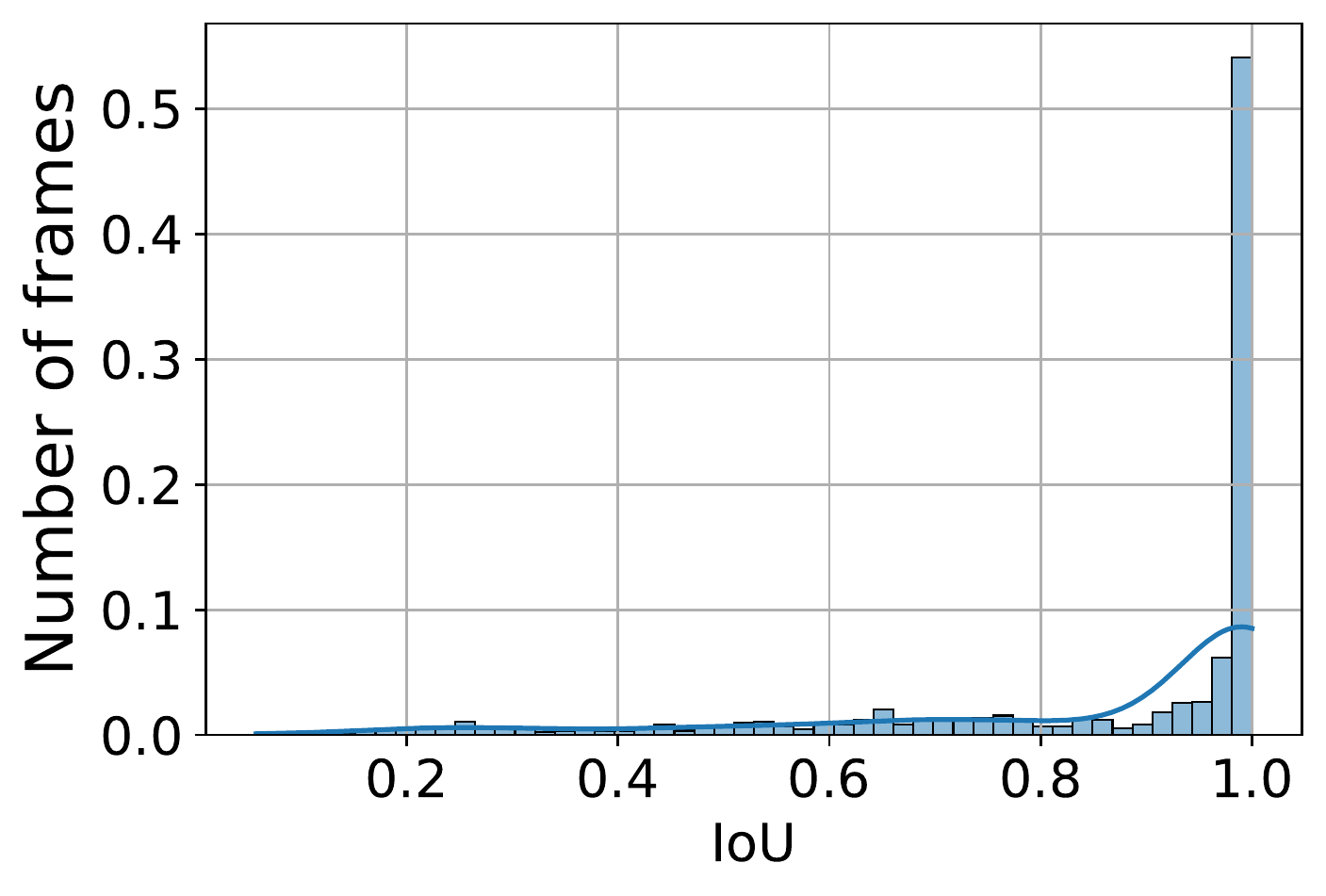}
         \caption{Defensive distillation}
         \label{fig:defended_iou_bim}
     \end{subfigure}
    \caption{BIM}
        \label{fig:bim_iou_bar}
\end{figure}

\begin{figure}[!tbp]
     \centering
     \begin{subfigure}[b]{0.49\linewidth}
         \centering
         \includegraphics[width=0.9\linewidth]{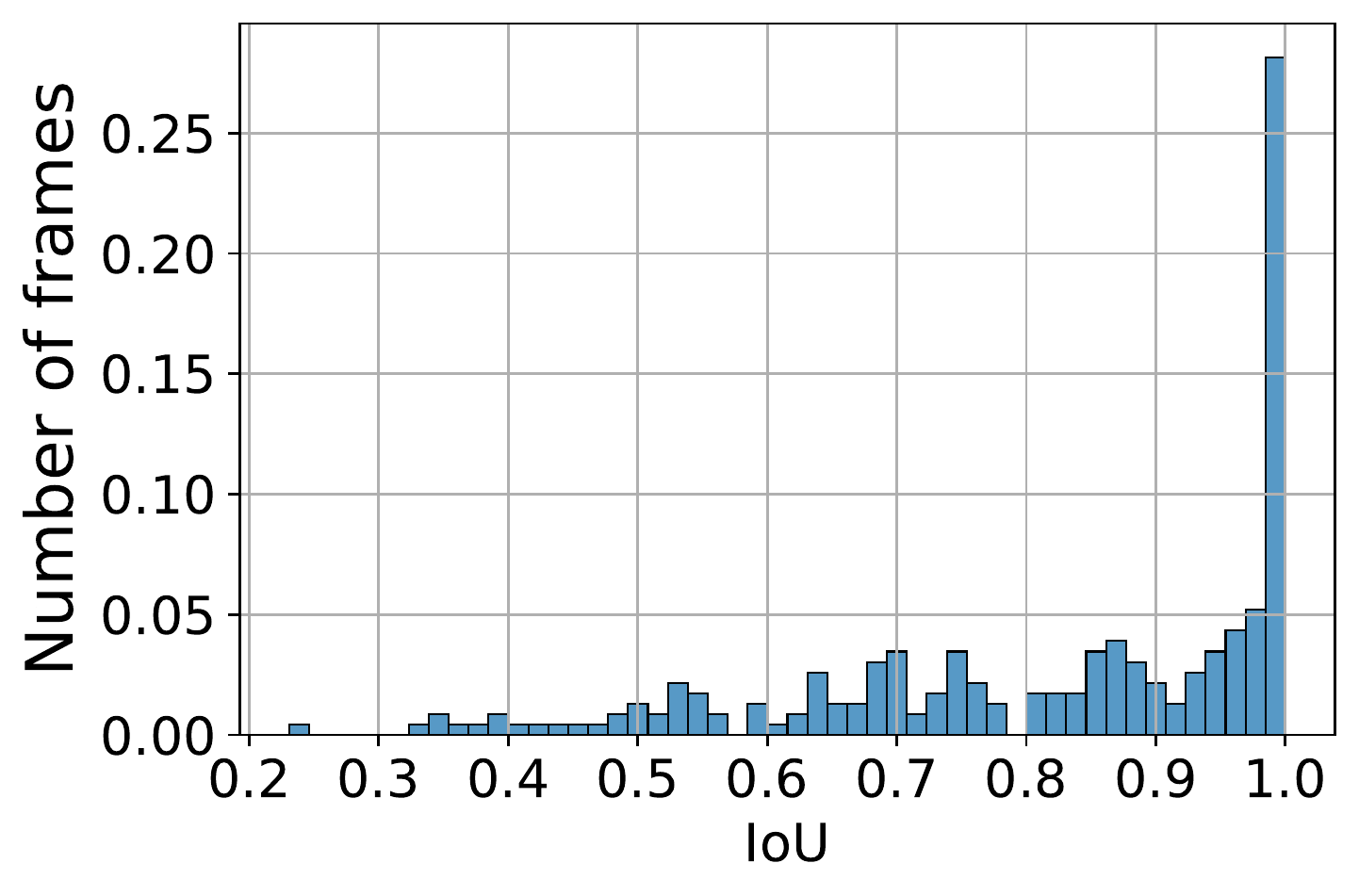}
         \caption{Undefended}
         \label{fig:undefended_iou_pgd}
     \end{subfigure}
     \hfill
     \begin{subfigure}[b]{0.49\linewidth}
         \centering
         \includegraphics[width=0.9\linewidth]{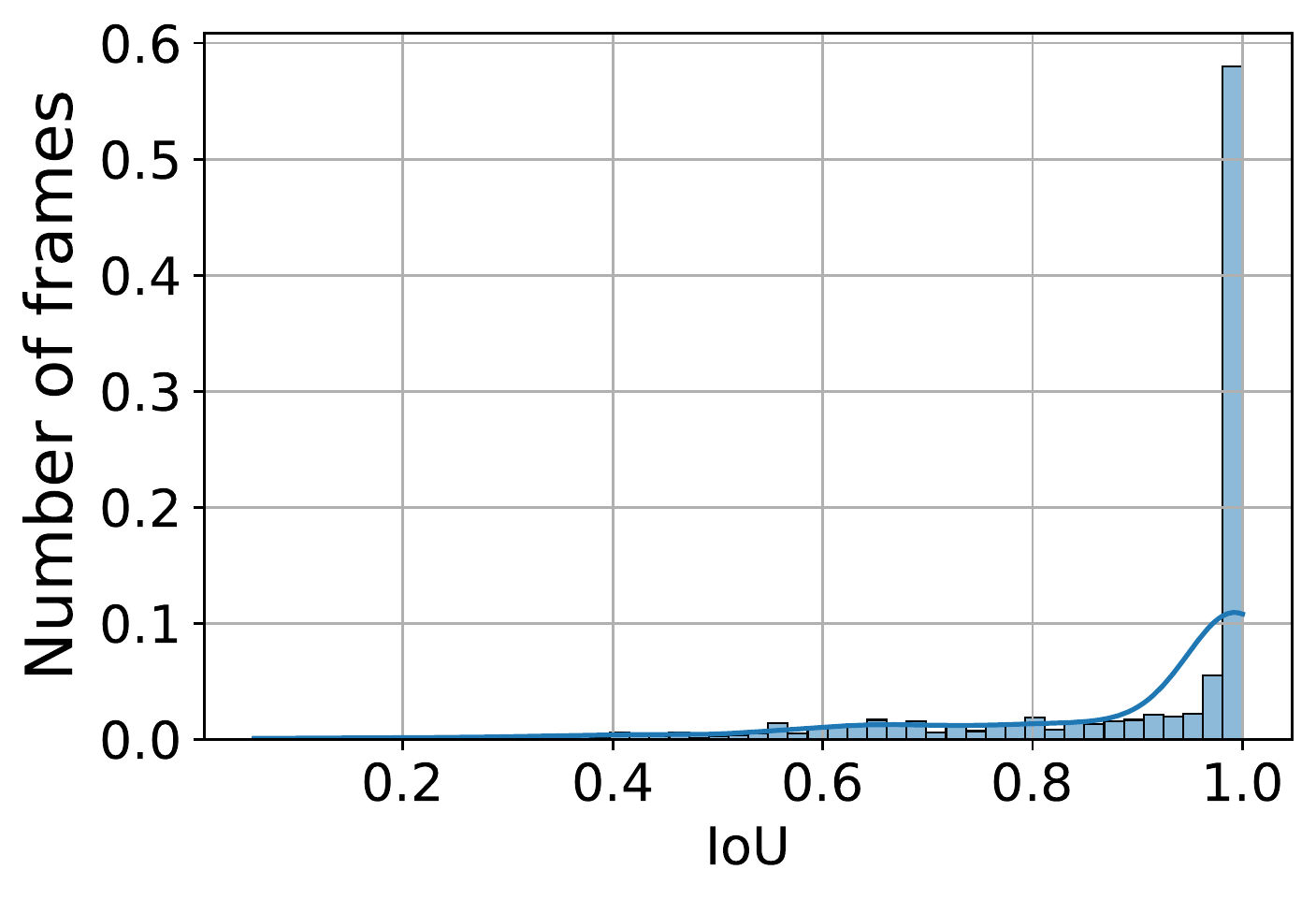}
         \caption{Defensive distillation}
         \label{fig:defended_iou_pgd}
     \end{subfigure}
    \caption{PGD}
        \label{fig:pgd_iou_bar}
\end{figure}

The trendlines are also shown in the figures, and the slopes of these trendlines show the defended model has a lower decrease rate when compared to the undefended model. The first obvious result is that the defensive distillation-based mitigation method can significantly reduce the vulnerabilities of  AI-based spectrum sensing models against adversarial attacks in cellular networks. 
Figure \ref{fig:fgsm_iou_bar}-\ref{fig:pgd_iou_bar} show the histogram plots of the IoU metric values for each attack of defended and undefended models. For the defended model, the IoU values are clustered around 1.0, which means the CNN model can correctly recognize the LTE, 5G, and noise signals. 

\section{Discussion}
\label{sec:discussion}
The results show that CNN-based spectrum sensing models are vulnerable to adversarial attacks. The IoU metric is also pretty much low, i.e., $~0.72$, under a lower power attack ($\epsilon$ equals 64, i.e., 0.25 of the maximum input value). Fortunately,  the mitigation method provides a better performance against higher-order adversarial attacks, and the IoU metric value goes up to $~0.87$. According to the results, adversarial attacks on CNN-based spectrum sensing models and the use of the  defensive adversarial mitigation method can be summarized:

\emph{\textit{Observation 1}}: The CNN-based spectrum sensing models are vulnerable to adversarial attacks.

\emph{\textit{Observation 2}}: There is a strong negative correlation between the attack power  \(\epsilon\) and the model performance.

\emph{\textit{Observation 3}}: The proposed mitigation method offers a better performance against adversarial attacks.
\
\emph{\textit{Observation 4}}: The defended model has a lower
decrease rate when compared to the undefended model.

\section{Conclusion}
\label{sec:conclusion}
With advanced computing and AI methods, new spectrum sensing approaches have been used  for better spectrum management in cellular networks. However, an AI-based model can be poisoned by adversarial attacks where malicious users inject fake training data with the aim of corrupting the learned model. This paper provides a vulnerability analysis of the spectrum sensing approach using AI-based models for identifying cellular network signals under adversarial attacks and training methods. It also presents the model performance with and without a mitigation method, i.e., defensive distillation, for adversarial attack. The results showed that the defensive distillation can defend the CNN-based spectrum sensing models against adversarial attacks in cellular networks. The CNN-based spectrum sensing model can identify the type of cellular signal, whether LTE, 5G, or noise. Simulation result show that the original CNN-based spectrum sensing model is significantly vulnerable to adversarial attacks, especially high-order ones. Fortunately, the proposed defensive distillation mitigation method can improve the performance of the spectrum sensing model and provide better results against higher-order adversarial attacks.

\bibliographystyle{IEEEtran}
\bibliography{references}

\begin{thebibliography}{10}
\providecommand{\url}[1]{#1}
\csname url@samestyle\endcsname
\providecommand{\newblock}{\relax}
\providecommand{\bibinfo}[2]{#2}
\providecommand{\BIBentrySTDinterwordspacing}{\spaceskip=0pt\relax}
\providecommand{\BIBentryALTinterwordstretchfactor}{4}
\providecommand{\BIBentryALTinterwordspacing}{\spaceskip=\fontdimen2\font plus
\BIBentryALTinterwordstretchfactor\fontdimen3\font minus
  \fontdimen4\font\relax}
\providecommand{\BIBforeignlanguage}[2]{{%
\expandafter\ifx\csname l@#1\endcsname\relax
\typeout{** WARNING: IEEEtran.bst: No hyphenation pattern has been}%
\typeout{** loaded for the language `#1'. Using the pattern for}%
\typeout{** the default language instead.}%
\else
\language=\csname l@#1\endcsname
\fi
#2}}
\providecommand{\BIBdecl}{\relax}
\BIBdecl

\bibitem{chataut2020massive}
R.~Chataut and R.~Akl, ``Massive {MIMO} systems for 5{G} and beyond
  networks—overview, recent trends, challenges, and future research
  direction,'' \emph{Sensors}, vol.~20, no.~10, p. 2753, 2020.

\bibitem{hu2018full}
F.~Hu, B.~Chen, and K.~Zhu, ``Full spectrum sharing in cognitive radio networks
  toward 5{G}: A survey,'' \emph{IEEE Access}, vol.~6, pp. 15\,754--15\,776,
  2018.

\bibitem{gupta2019progression}
M.~S. Gupta and K.~Kumar, ``Progression on spectrum sensing for cognitive radio
  networks: A survey, classification, challenges and future research issues,''
  \emph{Journal of Network and Computer Applications}, vol. 143, pp. 47--76,
  2019.

\bibitem{nasser2021spectrum}
A.~Nasser, H.~Al~Haj~Hassan, J.~Abou~Chaaya, A.~Mansour, and K.-C. Yao,
  ``Spectrum sensing for cognitive radio: Recent advances and future
  challenge,'' \emph{Sensors}, vol.~21, no.~7, p. 2408, 2021.

\bibitem{obite2021overview}
F.~Obite, A.~D. Usman, and E.~Okafor, ``An overview of deep reinforcement
  learning for spectrum sensing in cognitive radio networks,'' \emph{Digital
  Signal Processing}, vol. 113, p. 103014, 2021.

\bibitem{du2020spectrum}
K.~Du, P.~Wan, Y.~Wang, X.~Ai, and H.~Chen, ``Spectrum sensing method based on
  information geometry and deep neural network,'' \emph{Entropy}, vol.~22,
  no.~1, p.~94, 2020.

\bibitem{gao2019deep}
J.~Gao, X.~Yi, C.~Zhong, X.~Chen, and Z.~Zhang, ``Deep learning for spectrum
  sensing,'' \emph{IEEE Wireless Communications Letters}, vol.~8, no.~6, pp.
  1727--1730, 2019.

\bibitem{sarp2021use}
S.~Sarp, H.~Tang, and Y.~Zhao, ``Use of intelligent reflecting surfaces for and
  against wireless communication security,'' in \emph{2021 IEEE 4th 5G World
  Forum (5{GWF})}.\hskip 1em plus 0.5em minus 0.4em\relax IEEE, 2021, pp.
  374--377.

\bibitem{CATAK2022101626}
F.~O. Catak, M.~Kuzlu, E.~Catak, U.~Cali, and D.~Unal, ``Security concerns on
  machine learning solutions for 6g networks in mm{W}ave beam prediction,''
  \emph{Physical Communication}, vol.~52, p. 101626, 2022.

\bibitem{9527756}
E.~Catak, F.~O. Catak, and A.~Moldsvor, ``Adversarial machine learning security
  problems for 6{G}: mm{W}ave beam prediction use-case,'' in \emph{2021 IEEE
  International Black Sea Conference on Communications and Networking
  (BlackSeaCom)}, 2021, pp. 1--6.

\bibitem{wibawa2022homomorphic}
F.~Wibawa, F.~O. Catak, S.~Sarp, M.~Kuzlu, and U.~Cali, ``Homomorphic
  encryption and federated learning based privacy-preserving the {CNN}
  training: {COVID}-19 detection use-case,'' \emph{arXiv preprint
  arXiv:2204.07752}, 2022.

\bibitem{kuzlu2021role}
M.~Kuzlu, C.~Fair, and O.~Guler, ``Role of artificial intelligence in the
  internet of things ({IoT}) cybersecurity,'' \emph{Discover Internet of
  Things}, vol.~1, no.~1, pp. 1--14, 2021.

\bibitem{ma2009signal}
J.~Ma, G.~Y. Li, and B.~H. Juang, ``Signal processing in cognitive radio,''
  \emph{Proceedings of the IEEE}, vol.~97, no.~5, pp. 805--823, 2009.

\bibitem{zeng2010review}
Y.~Zeng, Y.-C. Liang, A.~T. Hoang, and R.~Zhang, ``A review on spectrum sensing
  for cognitive radio: challenges and solutions,'' \emph{EURASIP journal on
  advances in signal processing}, vol. 2010, pp. 1--15, 2010.

\bibitem{goodfellow2014explaining}
I.~J. Goodfellow, J.~Shlens, and C.~Szegedy, ``Explaining and harnessing
  adversarial examples,'' \emph{arXiv preprint arXiv:1412.6572}, 2014.

\bibitem{kurakin2016adversarial}
A.~Kurakin, I.~Goodfellow, and S.~Bengio, ``Adversarial machine learning at
  scale,'' \emph{arXiv preprint arXiv:1611.01236}, 2016.

\bibitem{kannan2018adversarial}
H.~Kannan, A.~Kurakin, and I.~Goodfellow, ``Adversarial logit pairing,''
  \emph{arXiv preprint arXiv:1803.06373}, 2018.

\bibitem{5537907}
Y.~LeCun, K.~Kavukcuoglu, and C.~Farabet, ``Convolutional networks and
  applications in vision,'' in \emph{Proceedings of 2010 IEEE International
  Symposium on Circuits and Systems}, 2010, pp. 253--256.

\bibitem{hinton2015distilling}
G.~Hinton, O.~Vinyals, and J.~Dean, ``Distilling the knowledge in a neural
  network,'' 2015.

\bibitem{papernot2016distillation}
N.~Papernot, P.~McDaniel, X.~Wu, S.~Jha, and A.~Swami, ``Distillation as a
  defense to adversarial perturbations against deep neural networks,'' 2016.

\bibitem{matlab_example_ss}
\BIBentryALTinterwordspacing
MathWorks, ``Spectrum sensing with deep learning to identify 5{G} and {LTE}
  signals.'' [Online]. Available:
  \url{https://www.mathworks.com/help/comm/ug/spectrum-sensing-with-deep-learning-to-identify-5g-and-lte-signals.html}
\BIBentrySTDinterwordspacing

\end{thebibliography}

\end{document}